\documentclass[aps,prb,superscriptaddress,twocolumn,longbibliography]{revtex4-2}
\usepackage{bbm}
\usepackage{graphicx}
\usepackage{dcolumn}
\usepackage{bm}
\usepackage{subfigure}
\usepackage{amsmath}
\usepackage{amssymb}

\usepackage{hyperref}
\hypersetup{
	colorlinks = true,
	citecolor  = magenta,  
	linkcolor  = magenta 
}

\usepackage{attachfile}

\newcommand{\bk}{\boldsymbol k}

\usepackage{times}

\begin{document}
\title{Topological superconductivity in two-dimensional $\pi$-junction Dirac semimetals}

\author{Yijie Mo}
\affiliation{Guangdong Provincial Key Laboratory of Magnetoelectric Physics and Devices, 
State Key Laboratory of Optoelectronic Materials and Technologies, School of Physics, Sun Yat-sen University, Guangzhou 510275, China}

\author{Xiao-Jiao Wang}
\affiliation{Guangdong Provincial Key Laboratory of Magnetoelectric Physics and Devices, State Key Laboratory of Optoelectronic Materials and Technologies, School of Physics, Sun Yat-sen University, Guangzhou 510275, China}

\author{Zhongbo Yan}
\email{yanzhb5@mail.sysu.edu.cn}
\affiliation{Guangdong Provincial Key Laboratory of Magnetoelectric Physics and Devices, 
State Key Laboratory of Optoelectronic Materials and Technologies, School of Physics, Sun Yat-sen University, Guangzhou 510275, China}

\date{\today}

\begin{abstract}
Odd-parity pairings offer a natural pathway for realizing topological superconductivity. When two identical even-parity superconductors form a 
$\pi$-junction, the metallic material sandwiched between them experiences an effective odd-parity pairing, facilitating the emergence of topological superconductivity in the intermediate region. In this work, we consider the intermediate material to be a two-dimensional spin-orbit-coupled 
Dirac semimetal. When the two superconductors are conventional 
$s$-wave superconductors, we find that a helical topological superconductor can be realized. 
This phase is characterized by the presence of a pair of helical Majorana edge states. Interestingly, when the superconductors are 
$s_{\pm}$-wave superconductors, we observe not only the helical topological superconductor but also an unconventional topological superconductor. The latter is distinguished by the existence of two pairs of helical Majorana edge states, despite the fact that the global topological invariants for this system take trivial values. By further applying an in-plane magnetic field, we demonstrate that second-order topological superconductors can be achieved. These phases host isolated Majorana corner modes as well as twofold Majorana corner modes. Our findings reveal that two-dimensional $\pi$-junction Dirac semimetals can support a rich variety of topological superconducting phases, offering a versatile platform for exploring exotic topological phenomena.
\end{abstract}

\maketitle

\section{Introduction}

Topological superconductors (TSCs) have attracted considerable attention 
over the past two decades due to their intriguing properties~\cite{qi2011,alicea2012new,leijnse2012introduction,Tanaka2012,stanescu2013majorana,
Beenakker2013,Elliott2015,Sato2016jpsj} and promising 
applications in topological quantum computation~\cite{nayak2008review,sarma2015majorana,Karzig2017MZM,Marra2022}. 
Although superconductors featuring intrinsic odd-parity pairings are naturally predisposed to 
exhibit topological nontriviality~\cite{read2000,kitaev2001unpaired,qi2009b}, the scarcity of such intrinsic 
odd-parity superconductors in nature poses a significant challenge to realizing this potential~\cite{Kallin2016review}. 
Consequently, the search for alternative approaches to engineer effective odd-parity 
pairings has emerged as a central focus in the quest to realize TSCs. 
A key breakthrough in this pursuit has been the identification of 
spin-orbit coupling (SOC) as a pivotal mechanism for enabling 
effective odd-parity pairings~\cite{fu2008}. Pioneering theoretical studies have demonstrated that 
the interplay of SOC, Zeeman (exchange) fields, and even-parity 
$s$-wave pairings can give rise to effective $s$-wave pairings~\cite{zhang2008px,sato2009non,sau2010,alicea2010,lutchyn2010,oreg2010helical}. This mechanism enables the realization of time-reversal-symmetry-breaking TSCs, including two-dimensional (2D) chiral 
TSCs hosting one-dimensional (1D) chiral Majorana edge modes~\cite{zhang2008px,sato2009non,sau2010,alicea2010}, as well as 1D TSCs 
featuring zero-dimensional (0D) Majorana zero modes at their ends\cite{lutchyn2010,oreg2010helical}. These groundbreaking 
theoretical insights have driven remarkable progress in the experimental exploration of TSCs~\cite{Mourik2012MZM,Nadj2014MZM,Sun2016Majorana,wang2018evidence,kong2019half,Fornieri2019,Ren2019Planar}.

In recent years, the emergence of the concept of higher-order topology has sparked 
renewed interest in the study of TSCs\cite{Benalcazar2017,Benalcazar2017prb,Song2017,Langbehn2017,Schindler2018}, 
as it expands the classification of TSCs and 
enriches platforms for hosting Majorana modes~\cite{Geier2018,Khalaf2018,Zhu2018hosc,Yan2018hosc,
Wang2018weak,Wang2018hosc,Liu2018hosc,Hsu2018hosc,Wu2019hosc,Yan2019hosca,Yan2019hoscb,Volpez2019SOTSC,Zhang2019hinge,Zhang2019hoscb,Pan2019SOTSC,
Zhu2019mixed,Ghorashi2019,Ahn2020hosc,Hsu2020hosc,Wu2020SOTSC,Majid2020hoscb,Franca2019SOTSC,
wu2020boundaryobstructedb,Laubscher2020mcm,Roy2020,Li2021bts,Luo2021hosc,Ghosh2021hierarchy,Qin2022SOTSC,
scammell2021intrinsic,Jahin2023HOTSC,Zhang2023interacting,Zhang2024TSC,Chatterjee2024,Bonetti2024HOTSC}, 
along with new mechanisms for their manipulation\cite{Zhu2018hosc,Zhang2020SOTSCb,Lapa2021braiding,Zhu2022sublattice1,Zhu2022sublattice2,Zhu2023sublattice3,Liu2024MCM}. 
With this expansion, TSCs are further categorized according to the codimension of the gapless 
Majorana modes on their boundaries. Specially, an $n$th-order TSC is defined by the 
presence of gapless Majorana modes with a codimension $d_{c}=n$~\cite{Schindler2018}. 
Within this classification scheme, conventional TSCs are regarded as first-order TSCs\cite{Trifunovic2019hoti}, 
as these gapless Majorana modes on their boundaries have a codimension of one. Interestingly, higher-order 
TSCs can be derived from first-order TSCs by selectively breaking their protecting symmetries. 
A notable example is the realization of a time-reversal-symmetry-breaking second-order TSC, 
achieved by applying an in-plane magnetic field to a helical
TSC (a first-order TSC protected by time-reversal symmetry~\cite{Haim2019review}) in two dimensions~\cite{Langbehn2017,Zhu2018hosc}.  
This example highlights that symmetry-protected first-order TSCs 
are not only intrinsically interesting but can also serve as parent states for the 
generation of higher-order TSCs.

Despite being an intriguing topological phase, the experimental realization of 
2D helical TSCs has remained elusive. 
To achieve this phase, two primary scenarios have been proposed. First, in the case of 
odd-parity pairing with an inversion-invariant Hamiltonian, the sole requirement for 
realization is the presence of an odd number of Kramers degenerate Fermi surfaces~\cite{fu2010odd,sato2010odd}. In contrast, for 
even-parity pairing, inversion symmetry must be broken, and the Fermi surfaces must 
be spin-split by SOC. Additionally, the even-parity pairing 
must exhibit sign changes across the Brillouin zone, ensuring that among the 
even-numbered segments of the Fermi surface, an odd number of them have positive 
pairing signs, while the remaining odd number have negative pairing signs~\cite{Qi2010invariant}.
For the first scenario, existing theoretical proposals have primarily focused on 
interaction-driven intrinsic odd-parity pairing in doped thin-film or monolayer 
topological insulators (TIs)~\cite{Wang2014TSC,Feng2023Timereversal},  bilayer systems with inversion-symmetric SOC~\cite{Deng2012TRITSC,Nakosai2012,Xu2024helical},
as well as bilayer Fermi gas immersed in a Bose-Einstein condensate~\cite{Wu2017TRITSC}. 
For the second scenario, representative theoretical proposals include heterostructures 
combining semiconductors with inversion-asymmetric Rashba SOC and unconventional 
superconductors~\cite{zhang2013kramers,Scheurer2015,Chen2018helical}, as well as thin films of iron-based superconductors with Dirac surface states~\cite{Zhang2021TSC} or bulk Dirac points~\cite{Majid2022}. 

\begin{figure}[t]
	\centering
	\includegraphics[width=0.48\textwidth]{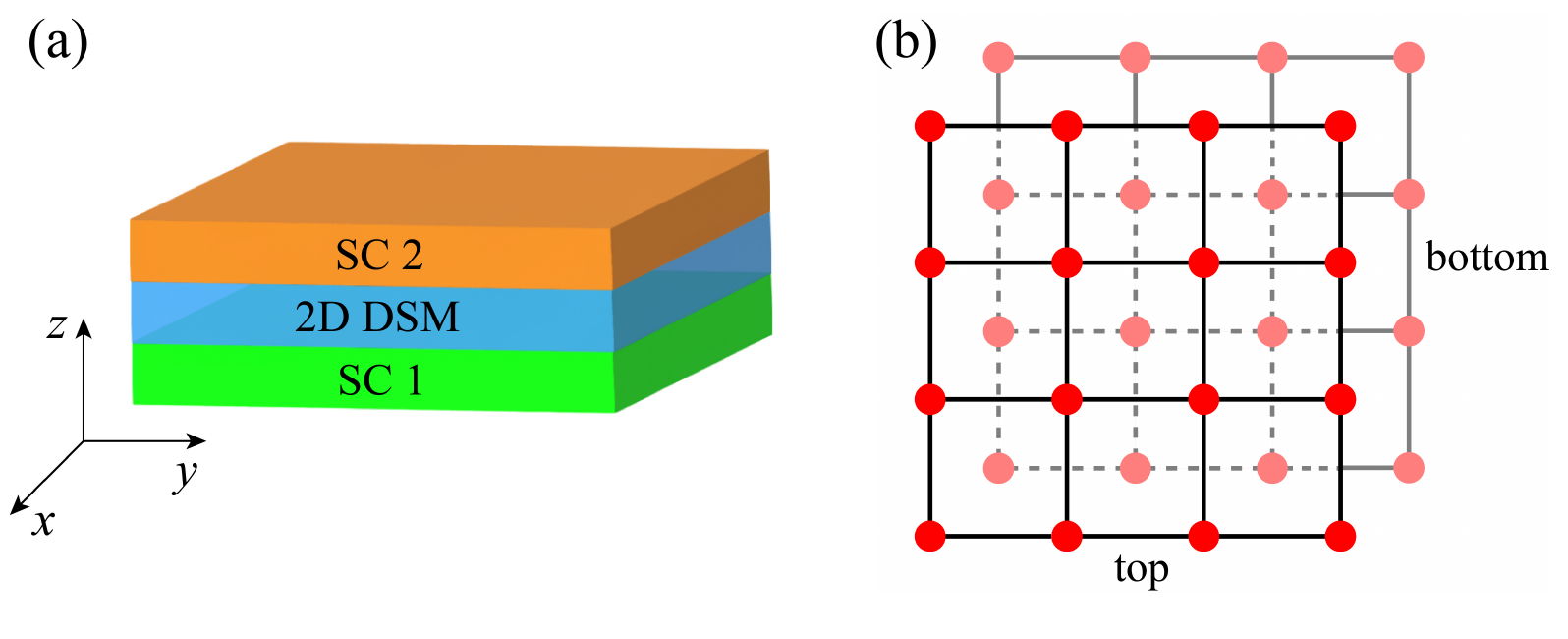}
	\caption{(a) Schematic of the $\pi$-junction. The outer two materials are superconductors with even-parity 
$s$-wave or $s_{\pm}$-wave pairing, and the intermediate material is a 2D DSM. The two superconductors exhibit a $\pi$ phase difference in 
their pairings. (b) Bilayer lattice structure of the 2D DSM. 
	}\label{fig1}
\end{figure}

Interestingly, in addition to intrinsic interaction-driven odd-parity pairing, effective odd-parity pairing can also be induced through the superconducting proximity effect in a $\pi$-junction structure, as illustrated in Fig.\ref{fig1}(a). Specifically, when two identical even-parity superconductors with a $\pi$ phase difference are coupled, the intermediate material between them experiences an effective odd-parity pairing. Previous studies have demonstrated that this approach can realize 2D helical TSCs in a bilayer system with inversion-symmetric 
SOC when it is sandwiched between two conventional $s$-wave superconductors with a $\pi$ phase difference~\cite{Volpez2019SOTSC}, 
or in a thin film of TI when the pairings for the Dirac surface states at the top and bottom surfaces
have a $\pi$ phase difference~\cite{Parhizgar2017,Pan2024MZM}. 
Inspired by these studies, we propose in this work to use a different class of topological materials 
as the intermediate layer: 2D spin-orbit-coupled Dirac semimetals (DSMs). These materials feature 
Dirac points in their band structure, which are stabilized by the combined action of 
time-reversal symmetry, inversion symmetry, and nonsymmorphic symmetries~\cite{Young2015DSM}. Furthermore, in addition to conventional 
$s$-wave superconductors, we also explore the use of unconventional superconductors with $s_{\pm}$-wave pairing, 
such as iron-based superconductors\cite{Stewart2011review,Hirschfeld2011}, to construct the $\pi$-junction.

Our findings can be summarized as follows. When the $\pi$-junction is formed by two $s$-wave 
superconductors, we observe that a helical TSC characterized by the presence of a pair of 
helical Majorana edge states can be achieved. 
When the $s$-wave superconductors are replaced by  $s_{\pm}$-wave superconductors, richer physics emerges. 
Specially,  in addition to the helical TSC, we find that the system 
can host an unconventional first-order TSC phase when the pairing nodes lie between 
two Fermi surface segments. Interestingly, this phase is characterized by the 
presence of two pairs of helical Majorana edge states, suggesting that it is $Z_{2}$ 
topologically trivial according to the tenfold-way classification\cite{Schnyder2008classification,kitaev2009periodic,Ryu2010topological}. 
The robustness of these two pairs of helical Majorana edge states can be attributed to 
weak topological invariants defined along high-symmetry lines within the Brillouin zone.
By applying an in-plane magnetic field to these TSC phases, we find that the helical 
Majorana edge states become gapped, leading to the emergence of Majorana zero modes 
at specific corners of the system. The increase in the codimension of the Majorana 
modes on the boundary suggests a transition from first-order TSCs to second-order TSCs. 
Interestingly, the second-order TSC phases in this system can host not only isolated 
Majorana corner modes but also twofold Majorana corner modes, a feature enabled by 
the existence of the unconventional TSC phase supporting two pairs of helical Majorana edge states.

The rest of this paper is organized as follows. In Sec.~\ref{II}, we introduce the setup and Hamiltonian, 
and analyze the salient properties in the normal-state band structure. 
In Sec.~\ref{III}, we explore first-order TSC phases 
that can emerge in this system. In Sec.~\ref{IV}, we investigate the transition from first-order TSCs 
to second-order TSCs driven by an in-plane magnetic field. Finally, in Sec.~\ref{V}, 
we present a discussion and conclusions.

\section{Setup and Hamiltonian}\label{II}

The setup concerned in this work is depicted in Fig.\ref{fig1}(a). Specially, a 2D DSM is 
sandwiched between two superconductors with a $\pi$ phase difference. The lattice structure 
of the 2D DSM is shown in Fig.\ref{fig1}(b). It comprises two layers, which are shifted 
relative to each other by $a(\frac{1}{2},\frac{1}{2})$, where $a$ represents the 
lattice constants of the square lattice. We assume that each layer of the DSM
inherits the same pairing type and phase from the fully-gapped superconductor in direct contact.
The effective tight-binding Hamiltonian describing the superconducting DSM is 
given by $H=\frac{1}{2}\sum_{\bk}\Psi_{\bk}^{\dag}\mathcal{H}_{\rm BdG}(\bk)\Psi_{\bk}$, where the basis 
function is defined as
 $\Psi^{\dag}_{k}=(c_{t,\uparrow,\bk}^{\dag}
,c_{t,\downarrow,\bk}^{\dag},c_{b,\uparrow,\bk}^{\dag},c_{b,\downarrow,\bk}^{\dag}, c_{t,\uparrow,-\bk}, c_{t,\downarrow,-\bk},c_{b,\uparrow,-\bk},c_{b,\downarrow,-\bk})$, and the Bogoliubov-de Gennes (BdG) Hamiltonian
is expressed as 
\begin{eqnarray}
	\mathcal{H}_{\rm BdG}(\mathbf{\bk})=\begin{pmatrix}
		\mathcal{H}_{\rm D}(\mathbf{\bk})-\mu & \Delta(\bk) \\
		\Delta^{\dag}(\bk) & -\mathcal{H}_{\rm D}^{T}(-\mathbf{\bk})+\mu
	\end{pmatrix}.\label{BdGH}
\end{eqnarray}  
Here, $\mathcal{H}_{\rm{D}}(\bk)$ describes the band structure of the normal-state DSM, $\mu$ is 
the chemical potential, and $\Delta(\bk)$ is the pairing matrix. The explicit form of $\mathcal{H}_{\rm{D}}(\bk)$ 
is~\cite{Young2015DSM} 
\begin{eqnarray}
\mathcal{H}_{\rm{D}}(\bk)&=&-2t(\cos k_{x}+\cos k_{y}) +4\eta\cos\frac{k_{x}}{2}\cos\frac{k_{y}}{2}\sigma_{x}\nonumber\\
&&+2\lambda_{\rm so}(\sin k_{x}\sigma_{z}s_{y}-\sin k_{y} \sigma_{z}s_{x}), \label{normal Hamiltonian}
\end{eqnarray}
and the pairing matrix $\Delta(\bk)$ is given by 
\begin{eqnarray}
\Delta(\bk)=-i\sigma_{z}s_{y}[\Delta_{0}+2\Delta_{1}(\cos k_{x}+\cos k_{y})].\label{pairing}
\end{eqnarray}
Here, the Pauli matrices $\{ \sigma_{i} \}$ and $\{ s_{i} \}$ act on the layer ($t, b$) and spin ($\uparrow, \downarrow$) degrees of freedom, respectively. In $\mathcal{H}_{\rm{D}}(\bk)$, the parameter $t$ represents the strength of the intralayer nearest-neighboring hopping, $\eta$ denotes 
the strength of the interlayer nearest-neighboring hopping, and $\lambda_{\rm so}$ denotes the strength of SOC. 
For $\Delta(\bk)$, $\Delta_{0}$ denotes the on-site $s$-wave pairing amplitude, and  $\Delta_{1}$ represents the amplitude of 
the intralayer nearest-neighboring pairing. When $\Delta_{0}$ is finite and $\Delta_{1}$ vanishes, 
the pairing order parameter describes a conventional $s$-wave superconductor.  When $4|\Delta_{1}|>|\Delta_{0}|$, 
the pairing order parameter has nodes in the Brillouin zone, and it describes an $s_{\pm}$-wave superconductor
if the pairing nodes lie between two Fermi surfaces of the superconductor. 
For notational simplicity, we set the lattice constant $a$ to unity and omit the identity
matrices in orbital and spin space throughout this work. 

It is straightforward to verify that the DSM Hamiltonian $\mathcal{H}_{\rm{D}}(\bk)$ possesses the following symmetries~\cite{Young2015DSM}:
time-reversal symmetry ($\mathcal{T}=is_{y}\mathcal{K}$ with $\mathcal{K}$ the 
complex conjugate operator), inversion symmetry ($\mathcal{P}=\sigma_{x}$), 
a glide mirror symmetry ($\{ \mathcal{M}_{z}|(\frac{1}{2},\frac{1}{2}) \}$ with $\mathcal{M}_{z}=i\sigma_{x}s_{z}$) 
and two screw symmetries ($\{ \mathcal{C}_{2x}|(\frac{1}{2},0) \}$ and $\{ \mathcal{C}_{2y}|(0,\frac{1}{2}) \}$ with 
$\mathcal{C}_{2x}=i\sigma_{x}s_{x}$ and $\mathcal{C}_{2y}=i\sigma_{x}s_{y}$). 
It is also straightforward to verify that the pairing is odd under inversion, satisfying $\mathcal{P}\Delta(\bk)\mathcal{P}^{T}=-\Delta(-\bk)$. 
This suggests that the pairing form given in Eq.(\ref{pairing}) describes an odd-parity spin-singlet pairing.

\begin{figure}[t]
	\centering
	\includegraphics[width=0.48\textwidth]{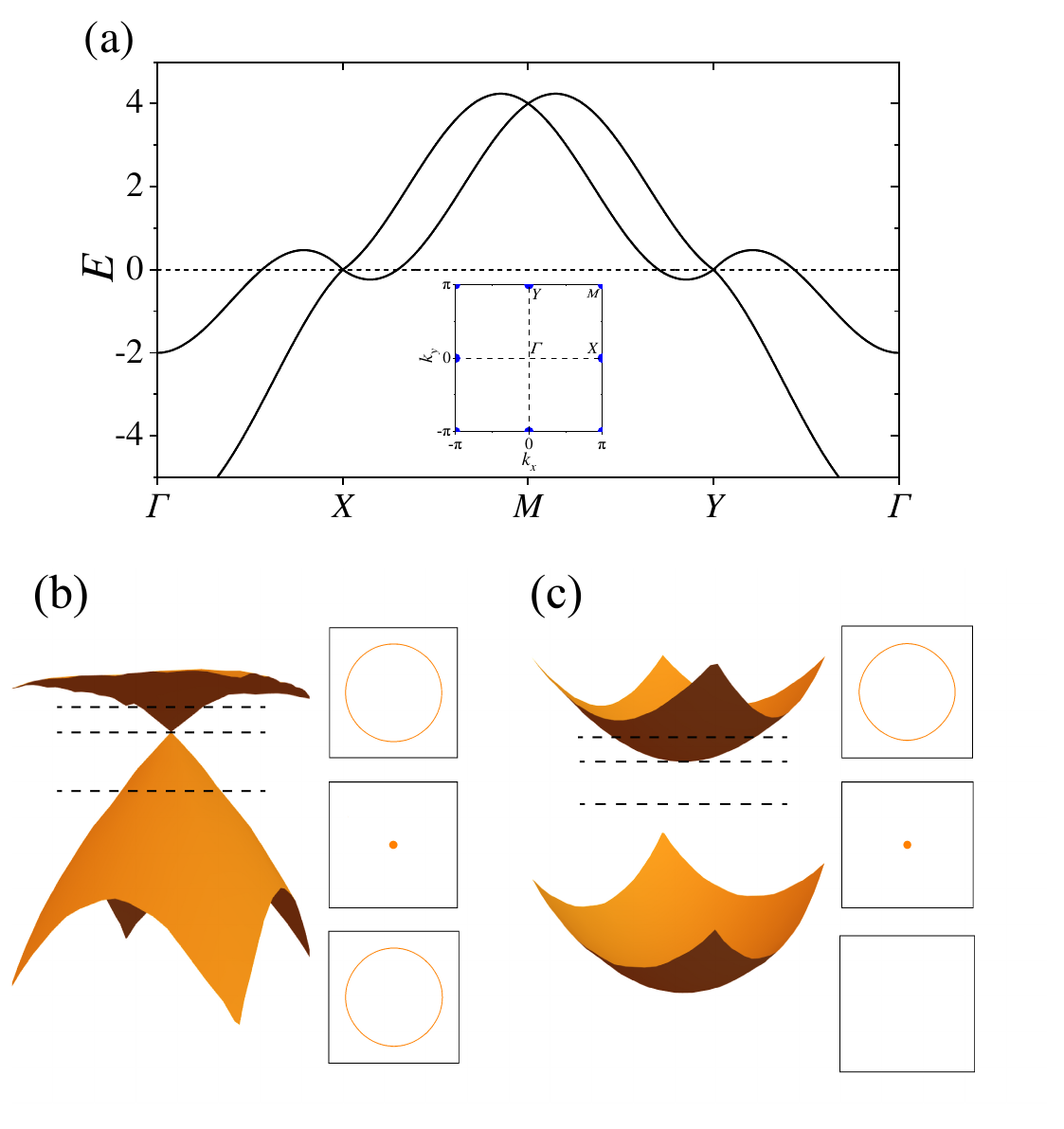}
	\caption{ (a) Representative band structure of the 2D DSM along high-symmetry lines. Dirac points are located at 
$\mathbf{X}=(\pi,0)$, $\mathbf{Y}=(0,\pi)$ and $\mathbf{M}=(\pi,\pi)$. The inset depicts the 2D Brillouin zone, 
highlighting the four time-reversal invariant momenta and the locations of Dirac points (blue dots). 
(b) Evolution of a DFSS with $\mu$. Left panel shows the Dirac-cone band structure near
 $\mathbf{M}$. Right panels (top to bottom) show Fermi surface structures around
  $\mathbf{M}$ for three different $\mu$ marked by dashed lines. 
(c) Evolution of a NFSS with $\mu$. Left panel shows the band structure near 
the $\boldsymbol{\Gamma}=(0,0)$ point. Right panels (top to bottom) show Fermi surface structures around $\boldsymbol{\Gamma}$
for three different $\mu$ marked by dashed lines. The values of parameters are: $t=1$, $\eta=0.5$, and $\lambda_{\rm so}=0.5$. 
	}\label{fig2}
\end{figure}

The symmetries in $\mathcal{H}_{\rm{D}}(\bk)$ guarantee the presence of four-fold degenerate Dirac points at the three time-reversal-invariant momenta: $\mathbf{X}=(\pi, 0)$, $\mathbf{Y}=(0, \pi)$ and $\mathbf{M}=(\pi, \pi)$. 
To see this more directly,  we write down the energy spectra of the DSM,  which read
\begin{eqnarray}
	\epsilon_{\pm}(\bk)=\epsilon_{0}(\bk)\pm \epsilon_{1}(\bk),
\end{eqnarray}
where $\epsilon_{0}(\bk)=-2t(\cos k_{x}+\cos k_{y})$ and 
$\epsilon_{1}(\bk)=\sqrt{(4\eta \cos \frac{k_{x}}{2} \cos \frac{k_{y}}{2})^2+4\lambda_{\rm so}^2(\sin ^2 k_{x}+\sin ^2 k_{y})}$.
Each band is doubly degenerate due to the simultaneous existence of time-reversal symmetry and inversion symmetry. 
It is evident that $\epsilon_{1}(\bk)$ vanishes at $\mathbf{X}$, $\mathbf{Y}$ and $\mathbf{M}$. 
This results in four-fold band degeneracies at these time-reversal-invariant momenta, corresponding to the Dirac points, 
as illustrated in Fig.\ref{fig2}(a). 

Since superconductivity arises from an instability of the Fermi surface, the properties of 
the Fermi surface significantly influence the resulting superconducting phase. 
In the current system, the Fermi surfaces exhibit markedly different behaviors 
depending on whether they enclose a Dirac point. Specifically, when a segment of 
the Fermi surface encloses a Dirac point, it cannot be continuously deformed to 
vanish unless it merges with other Fermi surface segments that also enclose Dirac points. 
In contrast, when a Fermi surface segment does not enclose a Dirac point, it can 
be continuously deformed to vanish by tuning the parameters of the Hamiltonian. 
An illustration of the different evolution behaviors of these two types of Fermi surface 
segments as a function of chemical potential is presented in Figs.\ref{fig2}(b) and \ref{fig2}(c). 
For clarity in the following discussion, 
we refer to Fermi surface segments that enclose a Dirac point as 
{\it Dirac Fermi surface segments} (DFSSs), and those that do not 
enclose a Dirac point as {\it normal Fermi surface segments} (NFSSs).

Several previous studies have demonstrated that the combination of this intriguing Dirac 
band structure with even-parity spin-singlet pairings can give rise to TSC phases with 
fascinating properties. Notably, since the DSM Hamiltonian also effectively describes 
the normal-state band structure of monolayer  FeSe, an iron-based superconductor, 
Qin {\it et al.} showed that when the pairing is an $s_{\pm}$-wave pairing and 
its nodes lie between two DFSSs, a time-reversal-invariant second-order TSC emerges, hosting Majorana 
Kramers pairs at the system's corners~\cite{Qin2022SOTSC}. In contrast, when the nodes 
lie between two NFSSs, the resulting superconducting phase is topologically trivial.
Furthermore, Mo {\it et al.} investigated $s$-wave superconductivity in this DSM and 
demonstrated that a perpendicular magnetic field can drive TSCs with large Chern numbers, 
as well as induce a nodal phase characterized by the coexistence of Bogoliubov Fermi surfaces 
and chiral Majorana edge states~\cite{Mo2024BFS}. The study of odd-parity superconductivity in this DSM can 
serve as a complementary direction to these studies, offering additional insights into the 
rich TSC phases achievable in such systems.

\section{First-order TSC phases}\label{III}

\subsection{General discussion of the topological characterization}
Before proceeding to determine the possible TSC phases arising from odd-parity pairings, 
we first rewrite the BdG Hamiltonian (\ref{BdGH}) in terms 
of Pauli matrices. This yields
\begin{eqnarray}
	\mathcal{H}_{\rm BdG}(\bk)&=&[-2t(\cos k_{x}+\cos k_{y})-\mu]\tau_{z} \nonumber \\
	&&+4\eta \cos \frac{k_{x}}{2} \cos \frac{k_{y}}{2} \tau_{z}\sigma_{x} \nonumber \\
	&&+2\lambda_{\rm so}(\sin k_{x}\tau_{z}\sigma_{z}s_{y}-\sin k_{y}\sigma_{z}s_{x}) \nonumber \\
	&&+[\Delta_{0}+2\Delta_{1}(\cos k_{x}+\cos k_{y})]\tau_{y}\sigma_{z}s_{y}. \label{PauliH}
\end{eqnarray}
Here, the Pauli matrices $\{ \tau_{i} \}$ act on the particle-hole space. For this BdG Hamiltonian, 
it possesses the following symmetries: time-reversal symmetry ($\mathcal{\tilde{T}}=is_{y}\mathcal{K}$), 
inversion symmetry ($\mathcal{\tilde{P}}=\tau_{z}\sigma_{x}$), particle-hole symmetry ($\Xi=\tau_{x}\mathcal{K}$), 
chiral symmetry ($\mathcal{S}=\tau_{x}s_{y}$), the glide mirror symmetry ($\{\mathcal{\tilde{M}}_{z}|(\frac{1}{2},\frac{1}{2}) \}$ with 
$\mathcal{\tilde{M}}_{z}=i\sigma_{x}s_{z}$), two screw symmetries ($\{ \mathcal{\tilde{C}}_{2x}|(\frac{1}{2},0) \}$ and 
$\{ \mathcal{\tilde{C}}_{2y}|(0,\frac{1}{2}) \}$ with $\mathcal{\tilde{C}}_{2x}=i\sigma_{x}s_{x}$ 
and $\mathcal{\tilde{C}}_{2y}=i\tau_{z}\sigma_{x}s_{y}$).

Since this 2D superconducting system possesses time-reversal symmetry, particle-hole symmetry and chiral symmetry, it belongs 
to the DIII class and follows a $Z_{2}$ classification according to the tenfold-way classification~\cite{Schnyder2008classification,kitaev2009periodic,Ryu2010topological}. 
Because the inversion symmetry is also preserved and the pairing is of odd-parity nature, the $Z_{2}$ invariant 
in the weak-pairing limit has a simple relation with the number of Kramers degenerate Fermi surface segments. Specially, 
the relation is given by~\cite{fu2010odd} 
\begin{eqnarray}
	(-1)^{\nu}=\prod_{i=1}^{4} (-1)^{N(\mathbf{K}_{i})},
	\label{Z2invariant}
\end{eqnarray}
where $\mathbf{K}_{i}\in\{\boldsymbol{\Gamma}, \mathbf{X},\mathbf{Y},\mathbf{M}\}$ 
represents a time-reversal-invariant momentum, 
and $N(\mathbf{K}_{i})$ denotes the number of Fermi surface segments 
that enclose $\mathbf{K}_{i}$. The simple formula (\ref{Z2invariant})
reveals that when the total number of Fermi surface segments is odd, the system 
is $Z_{2}$ nontrivial and realizes a helical 
TSC. This phase is characterized by a pair of helical Majorana edge states protected by 
time-reversal symmetry. 
In contrast, when the total number of Fermi surface segments is even, the system is 
$Z_{2}$ trivial, and  robust helical Majorana edge states are absent unless
additional topological classifications emerge from further symmetry protections~\cite{Yao2013classification,Chiu2013,Morimoto2013}.

In the current system, glide mirror symmetry introduces an additional topological classification. Since the mirror operator
 $\mathcal{\tilde{M}}_{z}$ commutes with the BdG Hamiltonian, 
$[\mathcal{\tilde{M}}_{z}, \mathcal{H}_{\rm BdG}(\bk)]=0$, the Hamiltonian $\mathcal{H}_{\rm BdG}(\bk)$
becomes block diagonal in the basis where $\mathcal{\tilde{M}}_{z}$ takes the diagonal form
$\text{diag}\{i I_{4\times4},-iI_{4\times4}\}$. Here $I_{4\times4}$ represents the $4\times4$ identity 
matrix. The block diagonal form of $\mathcal{H}_{\rm BdG}(\bk)$ can be expressed as 
$\mathcal{H}_{\rm BdG}(\bk)=\mathcal{H}_{+i}(\bk)\oplus \mathcal{H}_{-i}(\bk)$, where the subscript $+i$ ($-i$)
indicates that the block corresponds to the positive (negative) eigenvalue of the mirror operator. 
The explicit forms of the Hamiltonians for the two mirror sectors are given by
\begin{eqnarray}
	\mathcal{H}_{\pm i}(\bk)&=&[-2t(\cos k_{x}+\cos k_{y})-\mu]\tau_{z} 
	\nonumber \\
	&&+4\eta \cos \frac{k_{x}}{2} \cos \frac{k_{y}}{2} \tau_{z}\rho_{z} \nonumber \\
	&&-2\lambda_{\rm so}(\pm \sin k_{x}\tau_{z}\rho_{y}-\sin k_{y}\rho_{x}) \nonumber \\ &&\mp[\Delta_{0}+2\Delta_{1}(\cos k_{x}+\cos k_{y})]\tau_{y}\rho_{y}.
    \label{mirrorH}
\end{eqnarray}
Here, $\{ \rho_{i} \}$ denotes a new set of Pauli matrices that act on degrees of freedom 
arising from a combination of the layer and spin degrees of freedom. 
It is straightforward to verify that the Hamiltonians for both mirror sectors 
preserve the particle-hole symmetry. However, 
time-reversal symmetry and chiral symmetry are broken. Therefore, each mirror-sector Hamiltonian 
belongs to the D class and follows a $Z$ classification according to the tenfold-way classification~\cite{Schnyder2008classification,kitaev2009periodic,Ryu2010topological}.
The corresponding $Z$-valued topological invariant is the Chern number,  defined as~\cite{Ueno2013TCSC} 
\begin{eqnarray}
	C_{\pm i}=\frac{1}{2\pi}\sum_{n}\int_{BZ}\Omega_{\pm i}^{(n)}(\bk) d\bk,
\end{eqnarray}
where $\Omega_{\pm i}^{(n)}$ denotes the Berry curvature of the $n$th negative energy band in the $\pm i$ mirror-sector Hamiltonian, 
and the summation is taken over all negative energy bands.  Since the full Hamiltonian possesses time-reversal symmetry, 
the total Chern number, which is  the summation of the Chern numbers of the two mirror-sector Hamiltonians, is 
constrained by symmetry  to vanish identically. This also implies that the Chern numbers for 
the two mirror sectors satisfy the relation $C_{+i}=-C_{-i}$. Consequently, if one mirror sector hosts 
$n$ branches of chiral Majorana edge states, the other mirror sector must also support 
$n$ branches of chiral Majorana edge states, but with opposite chirality. Together, these form 
$n$ pairs of helical Majorana edge states. As a result, the relevant topological invariant in this 
context is the mirror Chern number, defined as\cite{Teo2008} 
\begin{eqnarray}
	C_{M}=\frac{1}{2}(C_{+i}-C_{-i}).
\end{eqnarray}
This invariant extends the topological classification from a $Z_{2}$ classification 
to a $Z$ classification and enables the presence of multiple pairs of robust helical 
Majorana edge modes. When the mirror Chern number $C_{M}$ is nonzero, the resulting 
phase is also referred to as an topological mirror superconductor~\cite{Zhang2013mirror}. Since the set of possible values 
for $\nu$ is a subset of those for $C_{M}$, we will primarily use $C_{M}$ to diagnose 
the topological classification of the resulting superconducting phases. \\

\subsection{First-order TSC phases in the $\pi$-junction formed by $s$-wave superconductors}

\begin{figure}[htp]
	\centering
	\includegraphics[width=0.48\textwidth]{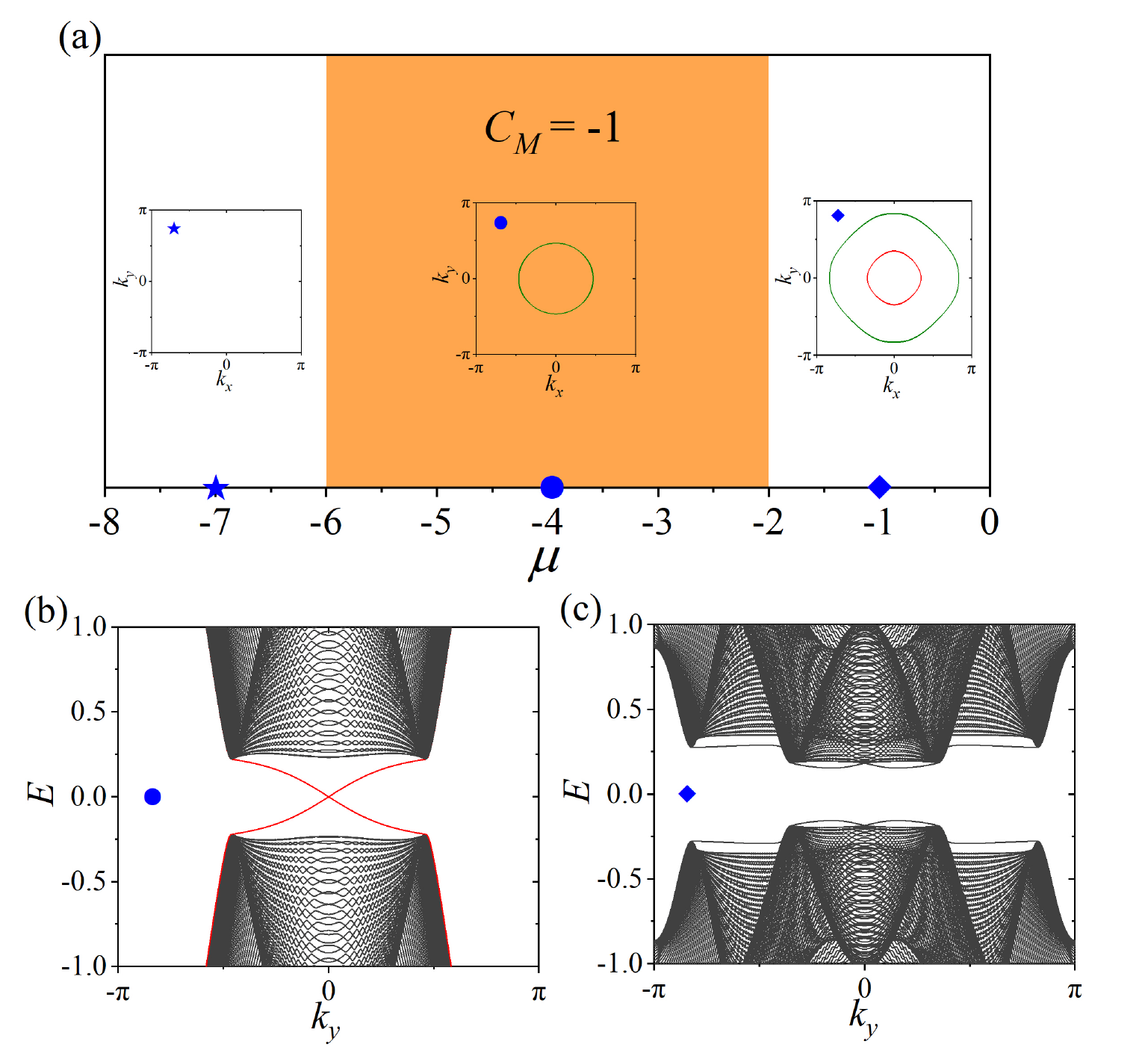}
	\caption{(a) Topological phase diagram determined by the mirror Chern number. Insets (left to right) 
show the Fermi surface structures corresponding to $\mu$ marked by star, dot and diamond, respectively. 
The two pairing parameters are set to $\Delta_{0}=0.1$ and $\Delta_{1}=0$. 
In (b) and (c), energy spectra for a system with periodic (open) boundary conditions 
in the $y$ ($x$) direction are shown. In (b), $\mu=-4$, a pair of helical Majorana modes (red lines) 
appears within the energy gap, consistent with the mirror Chern number $C_{M}=-1$.  
In (c), $\mu=-1$, no mid-gap states exist, consistent with $C_{M}=0$. In (b) and (c), the two pairing parameters 
are set to $\Delta_{0}=0.4$ and $\Delta_{1}=0$. A larger value of $\Delta_{0}$ is chosen to  more clearly highlight the edge states
within the energy gap. The values of other parameters are: $t=1$, $\eta=0.5$, and $\lambda_{\rm so}=0.5$.	}\label{fig3}
\end{figure}

We first focus on the case where $\Delta_{0}$ is finite and $\Delta_{1}$ vanishes. Since the change 
of first-order topology must be accompanied by the closing of bulk energy gap, we determine the 
gap-closing condition for this scenario. From Eq.(\ref{PauliH}), the energy spectra can be directly derived as
\begin{widetext}
\begin{eqnarray}
E_{\alpha\beta}(\bk)=\alpha\sqrt{\xi^{2}(\bk)+\eta^{2}(\bk)+\Lambda^{2}(\bk)+\Delta_{0}^{2}
+2\beta\sqrt{\xi^{2}(\bk)(\eta^{2}(\bk)+\Lambda^{2}(\bk))+\Delta_{0}^{2}\eta^{2}(\bk)}},
\end{eqnarray}
\end{widetext}
where $\alpha$ and $\beta$ take values in $\{+,-\}$, and we have introduced the short-hand notations: 
$\xi(\bk)=-2t(\cos k_{x}+\cos k_{y})-\mu$, $\eta(\bk)=4\eta\cos(k_{x}/2)\cos(k_{y}/2)$, 
and $\Lambda(\bk)=2\lambda_{\rm so}\sqrt{\sin^{2}k_{x}+\sin^{2}k_{y}}$. The energy gap can only close
at the four time-reversal invariant momenta where the parameter $\Lambda(\bk)$ vanishes identically. Based on this observation, 
the gap-closing condition can be readily derived as~\cite{sau2010}  
\begin{eqnarray}
\eta(\mathbf{K}_{i})=\pm\sqrt{\xi^{2}(\mathbf{K}_{i})+\Delta_{0}^{2}}.
\end{eqnarray}
Since $\eta(\mathbf{K}_{i})$ vanishes when $\mathbf{K}_{i}=(0,\pi)$, $(\pi,0)$ and $(\pi,\pi)$, 
the above equation can be satisfied only at the time-reversal invariant momentum $(0,0)$. At this point, the condition
is concretized  as
\begin{eqnarray}
4\eta=\pm\sqrt{(4t+\mu)^{2}+\Delta_{0}^{2}}.\label{criterion}
\end{eqnarray}
To simplify the discussion,  we restrict our analysis 
to the case where $\eta$ and $t$ are positive parameters.
Furthermore, we assume that $\mu$ is the only tuning parameter and that $4\eta>|\Delta_{0}|$, 
ensuring that gap closure can occur. Under these assumptions, Eq. (\ref{criterion}) shows that 
the bulk energy gap closes when the chemical potential $\mu$ takes the following critical values:
$\mu_{1}=-4t-\sqrt{16\eta^{2}-\Delta_{0}^{2}}$ and $\mu_{2}=-4t+\sqrt{16\eta^{2}-\Delta_{0}^{2}}$. 
To determine the topological phase diagram, we note that the system must be topologically trivial in the limit 
$\mu\rightarrow-\infty$ (where all bands in the normal state are empty) and in the limit $\mu\rightarrow+\infty$ (where all
bands in the normal state are occupied). 
By applying the adiabatic principle—which states that the first-order topology remains 
unchanged as long as the change of parameters does not induce the closing of the bulk energy gap—we 
conclude that the system can only be topologically nontrivial when $\mu_{1}<\mu<\mu_{2}$. 
Numerical calculations confirm this analysis, showing that the mirror Chern number $C_{M}$ equals $-1$
within this range and zero otherwise, as shown in Fig.\ref{fig3}(a).

In the weak-pairing limit, i.e., $|\Delta_{0}|\ll\eta$ and $|\Delta_{0}|\ll t$, we find that 
within the topological region, there is only a single Fermi surface segment. In contrast, in the trivial regions, 
the number of Fermi surface segments is even, as illustrated by the insets in Fig.\ref{fig3}(a). From Eq.(\ref{Z2invariant}), 
we know that the $Z_{2}$ invariant $\nu$ equals $1$ when a single Fermi surface segment exists, 
and $0$ when the number of Fermi surface segments is even. This demonstrates the consistency between 
the mirror Chern number $C_{M}$ and the $Z_{2}$ invariant $\nu$. 

In Figs. \ref{fig3}(b) and \ref{fig3}(c), we present the energy spectra of the system under open boundary conditions in the 
$x$ direction and periodic boundary conditions in the $y$ direction. Fig. \ref{fig3}(b) corresponds to the system being 
in the topological region, where we observe a pair of helical Majorana edge states traversing the bulk energy gap. 
In contrast, Fig. \ref{fig3}(c) illustrates the trivial region, where no edge states are present within the bulk 
energy gap. The presence or absence of helical Majorana edge states, as well as their number, is fully consistent 
with the topological invariant.

To conclude this section, we find that when a $\pi$-junction is formed between two 
$s$-wave superconductors, only one first-order TSC phase exists. This phase is 
characterized by the $Z_{2}$ invariant $\nu=1$ and the mirror Chern number $C_{M}=-1$, 
and is manifested by the presence of a pair of helical Majorana edge states localized at the boundary.
\\

\subsection{First-order TSC phases in the $\pi$-junction formed by $s_{\pm}$-wave superconductors}

We now investigate the TSC phases that can arise in a 
$\pi$-junction formed by two $s_{\pm}$-wave superconductors. The key distinction 
between the odd-parity pairing induced by $s$-wave superconductors and that induced by 
$s_{\pm}$-wave superconductors lies in the presence or absence of pairing nodes. 
As previously mentioned, Qin {\it et al.} demonstrated that when the pairing nodes of an even-parity 
$s_{\pm}$-wave pairing are located between two DFSSs, the system realizes a time-reversal-invariant second-order 
TSC phase characterized by the presence of Majorana corner modes~\cite{Qin2022SOTSC}. In contrast, when the pairing nodes 
lie between two NFSSs, the system becomes a trivial superconductor. 
This insight motivates us to explore analogous configurations in the present context 
to determine whether similar or distinct behaviors emerge.

To construct the topological phase diagram, we first determine the gap-closing conditions. We assume that the chemical potential
$\mu$ is the only variable,  while the parameters $t$, $\eta$,
$\lambda_{\rm so}$ and $\Delta_{1}$ are held fixed as positive constants. 
Under these assumptions, we find that the energy gap closes under the following conditions:

(I)  $|\Delta_{0}|>4\Delta_{1}$ (no pairing nodes; energy gap closes at high symmetry points). 

In this regime,  the energy gap closes only at the $\boldsymbol{\Gamma}$ point
when $\mu$ takes one of the following two values: 
\begin{eqnarray}
\mu_{1}^{\prime}&=&-4t-\sqrt{16\eta^{2}-(\Delta_{0}+4\Delta_{1})^{2}}, \nonumber\\
\mu_{2}^{\prime}&=&-4t+\sqrt{16\eta^{2}-(\Delta_{0}+4\Delta_{1})^{2}}.\label{condition2}
\end{eqnarray} 

(II)  $0<\Delta_{0}<4\Delta_{1}$ (pairing nodes enclose the $\boldsymbol{M}$ point; energy gap closes at 
points of high symmetry lines). 

For this case, the energy gap can close not only at the $\boldsymbol{\Gamma}$ point (with 
$\mu$ given by Eq.(\ref{condition2})), but also at the Brillouin zone boundary. 
To illustrate this, we note that the parameter $\eta(\bk)$ vanishes when $k_{x}=\pi$ or $k_{y}=\pi$. 
Focusing on the line $k_{x}=\pi$, the energy spectrum simplifies to $E_{\pm,\pm}(\pi,k_{y})=\pm\sqrt{\xi_{\pm}^{2}(k_{y})+\Delta^{2}(k_{y})}$, where 
$\xi_{\pm}(k_{y})=-2t(\cos k_{y}-1)-\mu\pm2\lambda_{\rm so}\sin k_{y}$ and $\Delta(k_{y})=\Delta_{0}+2\Delta_{1}(\cos k_{y}-1)$. 
The energy gap closes at $k_{y}=\pm\arccos\left(\frac{2\Delta_{1}-\Delta_{0}}{2\Delta_{1}}\right)$
when $\mu$ takes the values: 
\begin{eqnarray}
\mu_{3}^{\prime}&=&\frac{t\Delta_{0}}{\Delta_{1}}-2\lambda_{\rm so}\sqrt{1-\left(\frac{\Delta_{0}-2\Delta_{1}}{2\Delta_{1}}\right)^{2}},\nonumber\\
\mu_{4}^{\prime}&=&\frac{t\Delta_{0}}{\Delta_{1}}
+2\lambda_{\rm so}\sqrt{1-\left(\frac{\Delta_{0}-2\Delta_{1}}{2\Delta_{1}}\right)^{2}}.\label{condition3}
\end{eqnarray}
Similarly, along the line $k_{y}=\pi$, the energy gap closes at 
$k_{x}=\pm\arccos\left(\frac{2\Delta_{1}-\Delta_{0}}{2\Delta_{1}}\right)$ for the same values of $\mu$ given in Eq.(\ref{condition3}). 

(III)  $|\Delta_{0}|<4\Delta_{1}$ (pairing nodes present; energy gap closes at generic points).

When pairing nodes are present, the energy gap can also close at generic points in the Brillouin zone that simultaneously satisfy
\begin{eqnarray}
&\Delta_{0}+2\Delta_{1}(\cos k_{x}+\cos k_{y})=0,\nonumber\\
&\xi(\bk)\pm\sqrt{\eta^{2}(\bk)+\Lambda^{2}(\bk)}=0.
\end{eqnarray}
Since these solutions correspond to generic points in the Brillouin zone, 
analytical expressions for the critical values of $\mu$ cannot be obtained. 
Nevertheless, the gap-closing conditions  can be determined numerically. 

By numerically calculating the mirror Chern number $C_{M}$, we find that 
it changes value only when the energy gap closes at the $\boldsymbol{\Gamma}$
point.  This behavior is identical to the previous case where the $\pi$-junction is formed by 
$s$-wave superconductors. Specially,  we obtain $C_{M}=-1$ for $\mu_{1}^{\prime}<\mu<\mu_{2}^{\prime}$, corresponding to a regime with a single Fermi surface segment in the weak-pairing limit. Outside this range,
$C_{M}=0$, and the number of Fermi surface segments is even. 
  
\begin{figure}[t]
	\centering
	\includegraphics[width=0.52\textwidth]{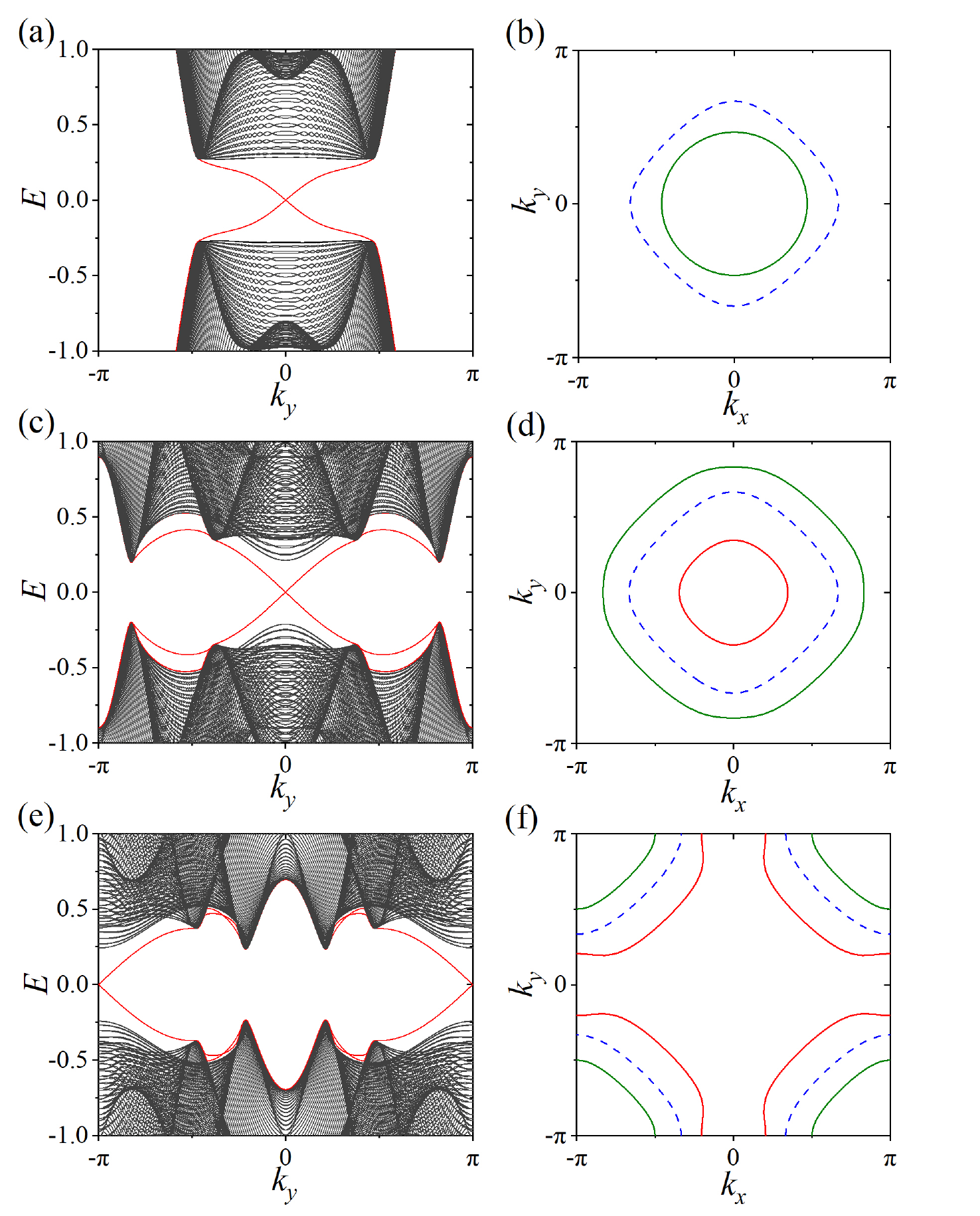}
	\caption{Left panels: Energy spectra for a system with periodic (open) boundary conditions in the $y$ ($x$) direction.  
    Right panels: Fermi surface structure (solid lines) and pairing nodes (dash lines) correspond to the left panels. 
    From top to bottom, parameters $\{\mu, \Delta_0, \Delta_{1}\}$ are set as $\{-4,-0.4,0.4\}$, $\{-1,-0.4,0.4\}$, and $\{1,0.4,0.4\}$, respectively.  The values of other parameters are: $t=1$, $\eta=0.5$, and $\lambda_{\rm so}=0.5$. 
	}\label{fig4}
\end{figure}

In the region where only a Fermi surface segment exists and $C_{M}=-1$, 
a pair of helical Majorana edge states is also observed to appear within the bulk energy gap,  
as shown in Figs.\ref{fig4}(a) and \ref{fig4}(b). 
Interestingly, when the system has two Fermi surface segments and the pairing nodes lie between them, 
we find two pairs of robust helical Majorana edge states, even though both 
the mirror Chern number $C_{M}$ and the $Z_{2}$ invariant are zero.  
Notably, these helical Majorana edge states emerge regardless of whether the two Fermi surface segments are  
DFSSs or NFSSs, as illustrated in Figs.\ref{fig4}(c)-\ref{fig4}(f). 
This behavior is markedly distinct from the even-parity pairing case, where nontrivial 
topology emerges exclusively when the pairing nodes lie between two DFSSs~\cite{Qin2022SOTSC}. 

How can we understand the presence of two pairs of robust helical Majorana edge states when 
the mirror Chern number $C_{M}$ is zero? We find that this phenomenon can be explained using 
weak topological invariants defined along the high-symmetry lines of the Brillouin zone~\cite{Feng2023Timereversal}. 
Specifically, we note that while each mirror-sector Hamiltonian belongs to the D class, the reduced 1D Hamiltonian along these high-symmetry lines exhibits an emergent chiral symmetry, placing it in the BDI class~\cite{Schnyder2008classification,kitaev2009periodic,Ryu2010topological}. The BDI class, characterized by higher symmetry, follows a
$Z$ classification in 1D. To illustrate this, consider the two mirror-sector Hamiltonians along the line $k_{y}=0$, 
\begin{eqnarray}
\mathcal{H}_{\pm i}(k_{x},0)&=&[-2t(\cos k_{x}+1)-\mu]\tau_{z} 
	\nonumber \\
	&&+4\eta \cos \frac{k_{x}}{2} \tau_{z}\rho_{z}\mp2\lambda_{\rm so}\sin k_{x}\tau_{z}\rho_{y} \nonumber \\ 
    &&\mp[\Delta_{0}+2\Delta_{1}(\cos k_{x}+1)]\tau_{y}\rho_{y}.
\end{eqnarray}
It is straightforward to verify that these two reduced 1D Hamiltonians 
possess chiral symmetry, with the chiral operator given by $\mathcal{S}_{0}=\tau_{x}$. 
Similarly, the two mirror-sector Hamiltonians along the line  $k_{y}=\pi$ take the form  
\begin{eqnarray}
\mathcal{H}_{\pm i}(k_{x},\pi)&=&[-2t(\cos k_{x}-1)-\mu]\tau_{z}\mp2\lambda_{\rm so}\sin k_{x}\tau_{z}\rho_{y}
	\nonumber \\
    &&\mp[\Delta_{0}+2\Delta_{1}(\cos k_{x}-1)]\tau_{y}\rho_{y},
\end{eqnarray}
which also possess chiral symmetry, with $\mathcal{S}_{\pi}=\tau_{x}$. The band topology of these 1D Hamiltonians is characterized 
by a winding number, defined as~\cite{Ryu2010topological} 
\begin{eqnarray}
	W_{\pm i}^{0/\pi}&=&\frac{1}{4\pi i}\oint dk_{x} {\rm{Tr}} [\mathcal{S}_{0/\pi}H_{\pm i}^{-1}(k_{x}, k_{y}=0/\pi)  \nonumber \\
	&&\times \partial_{k_{x}}H_{\pm i}(k_{x}, k_{y}=0/\pi)].
	\label{mirrorwn}
\end{eqnarray}
Here, $W_{\pm i}^{0}$ and $W_{\pm i}^{\pi}$ quantify the number of Majorana zero modes per boundary at $k_{y}=0$ and $k_{y}=\pi$, respectively. 
Through numerical calculations, we find that the mirror-graded winding number $W_{\pm i}^{0}$ equals $-2$ for the 
configuration of pairing nodes and Fermi surfaces shown in Fig.\ref{fig4}(d). The combination of $C_{\pm i}=0$ and $W_{\pm i}^{0}=-2$
implies the presence of a pair of helical Majorana edge states in each mirror sector. Furthermore, $W_{\pm i}^{0}=-2$ indicates 
the presence of four Majorana zero modes per boundary when $k_{y}=0$, corresponding to an eightfold crossing of the edge states' spectrum
at  $E=0$ and $k_{y}=0$, as shown in Fig.\ref{fig4}(c). Similarly, the combination of $C_{\pm i}=0$ and $W_{\pm i}^{\pi}=2$
also suggests the presence of a pair of helical Majorana edge states in each mirror sector. Additionally, $W_{\pm i}^{\pi}=2$ indicates that the edge states' spectrum crosses at $E=0$ occurs when $k_{y}=\pi$, as shown in Fig.\ref{fig4}(e). 

\begin{figure}[htp]
	\centering
	\includegraphics[width=0.49\textwidth]{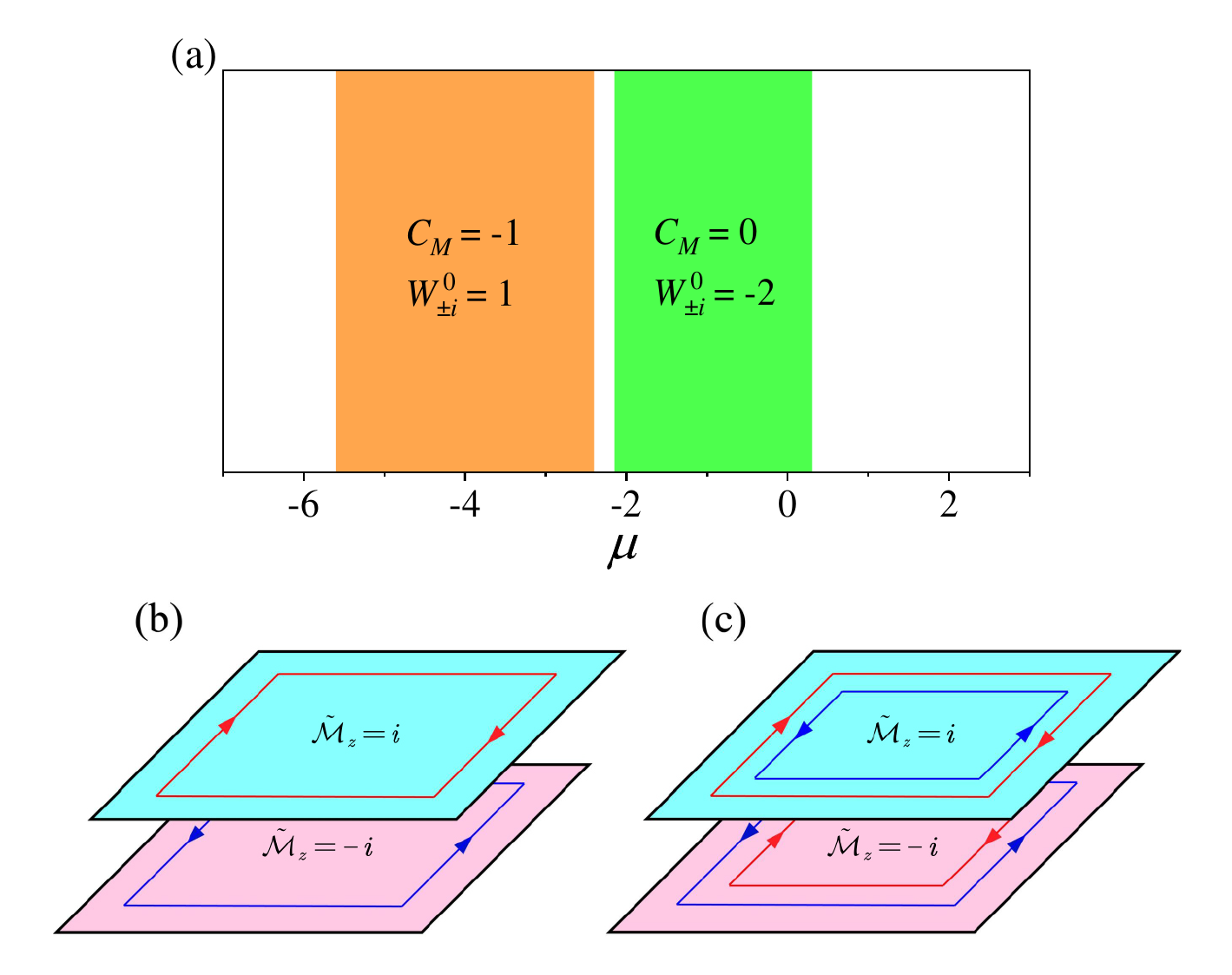}
	\caption{(a) Representative topological phase diagram determined by 
the mirror Chern number $C_{M}$ and the weak topological invariants $W_{\pm i}^{0}$. 
The parameter values are: $t=1$, $\eta=0.5$, 
and $\lambda_{\rm so}=0.5$,  $\Delta_{0}=-0.4$ and $\Delta_{1}=0.4$. For this set of parameters, 
$W_{\pm i}^{\pi}$ vanish identically and is therefore not considered.  
(b) Schematic of the edge states for the TSC phase 
characterized by $C_{M}=-1$ and $W_{\pm i}^{0}=1$. (c) Schematic of the edge states for the TSC phase 
characterized by $C_{M}=0$ and $W_{\pm i}^{0}=-2$.  
	}\label{fig5}
\end{figure} 

A combination of $C_{M}$ and $W_{\pm i}^{0/\pi}$ provides a comprehensive understanding of the topology of the resulting superconducting phases. A representative topological phase diagram based on these invariants is shown in Fig.~\ref{fig5}(a). 
Within the topological phase diagram, the TSC phase characterized by 
$C_{M}=-1$ and $W_{\pm i}^{0}=1$ hosts one chiral Majorana edge state in each mirror sector. Furthermore, 
the chiral Majorana edge states in the two mirror sectors exhibit opposite chirality, as illustrated in 
Fig.~\ref{fig5}(b). On the other hand, the TSC phase characterized by 
$C_{M}=0$ and $W_{\pm i}^{0}=-2$ (or $W_{\pm i}^{\pi}=2$) hosts one pair of helical Majorana edge state
in each mirror sector, as illustrated in Fig.~\ref{fig5}(c).

\section{Second-order TSC phases} \label{IV}

As previously mentioned, a transition from time-reversal-invariant first-order 
TSC phases to a time-reversal-symmetry-broken second-order TSC phase can be 
induced by applying an in-plane magnetic field. Previous studies have demonstrated 
that gapping out one pair of helical Majorana edge states can lead to the emergence 
of isolated Majorana zero modes at specific corners of an open-boundary system~\cite{Langbehn2017,Zhu2018hosc}. 
In our system, a TSC phase with two pairs of helical Majorana edge states exists, which suggests 
the possibility of observing twofold Majorana zero modes at the corners~\cite{Laubscher2020mcm,Yin2025}.
To investigate this, we introduce an in-plane magnetic field, which induces a Zeeman field 
described by $\mathcal{H}_{Z}=M_{x}\tau_{z}s_{x}+M_{y}s_{y}$ into the BdG Hamiltonian (\ref{PauliH}). 
This Zeeman field breaks both time-reversal symmetry and glide mirror symmetry, resulting in a 
coupling between the chiral or helical Majorana edge states in the two mirror sectors.
 
\begin{figure}[htp]
	\centering
	\includegraphics[width=0.48\textwidth]{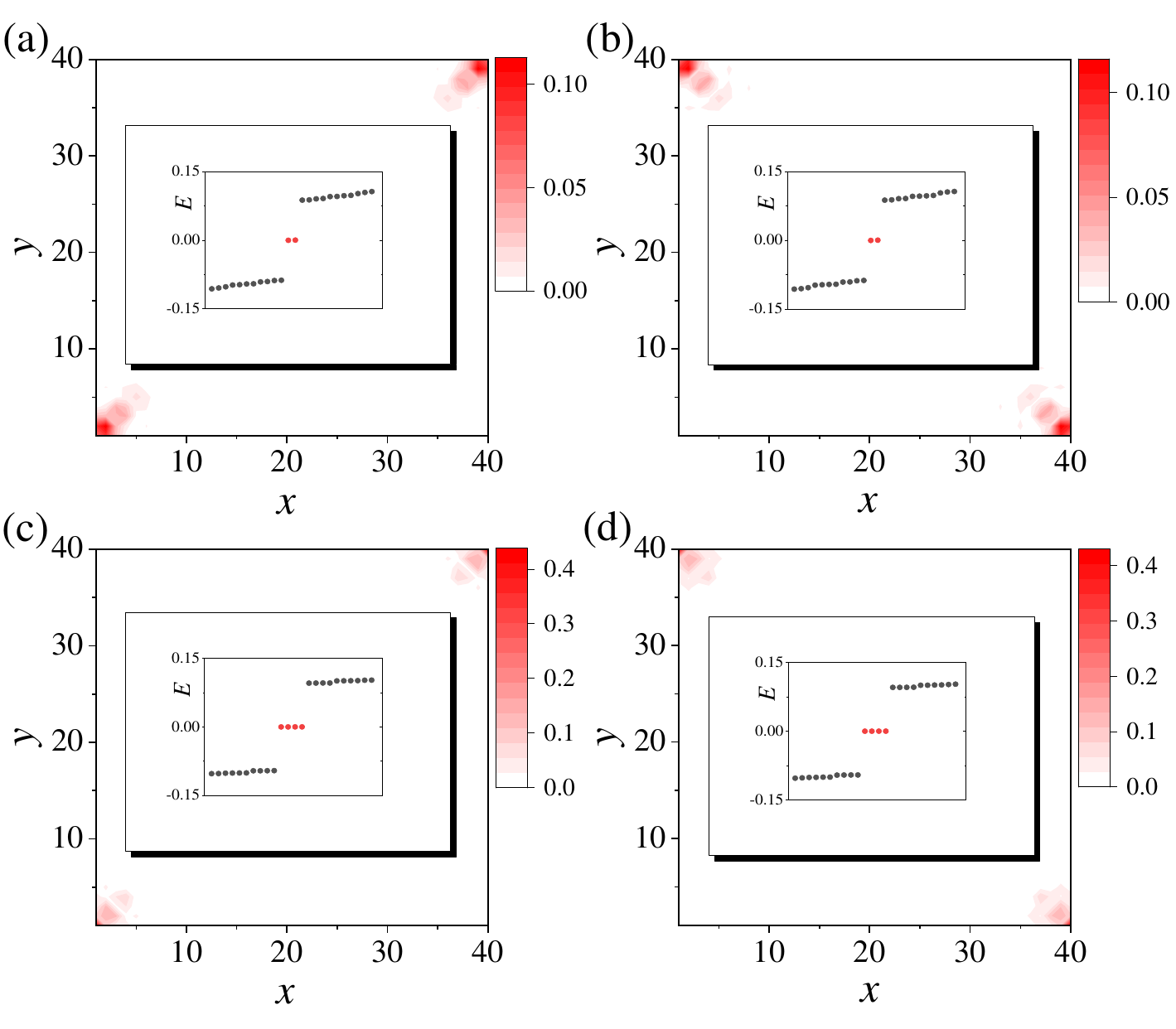}
	\caption{Probability density profiles of Majorana corner modes in an open boundary 
system of size $N_{x}\times N_{y}=40\times40$. Insets show a few eigenenergies closest to $E=0$. 
Parameters $\{\mu,\Delta_{0},\Delta_{1}\}$ are set as $\{-4,-0.4,0.4\}$ and 
$\{1,0.4,0.4\}$ for the top and bottom panels, respectively. Parameters 
$\{M_{x},M_{y}\}$ are chosen as $\{0.1,0.1\}$  and $\{0.1,-0.1\}$ for the left and right panels, respectively. 
The top (bottom) two panels show the presence of isolated (twofold) Majorana corner modes at two of the four corners. 
The values of other parameters are: $t=1$, $\eta=0.5$, and $\lambda_{\rm so}=0.5$.
	}\label{fig6}
\end{figure}

Specifically, we consider a system with open boundary conditions in both the 
$x$ and $y$ directions, subjected to a Zeeman field aligned along either the 
diagonal or anti-diagonal direction. By computing the energy spectrum, we observe 
that the helical Majorana edge states are gapped, leaving only two or four 
zero-energy states within the spectral gap. As expected, the number of these 
zero-energy states is directly tied to the number of helical Majorana edge state. 
These mid-gap states are identified as Majorana zero modes, and their wave functions 
are strongly localized at two inversion-related corners. Notably, the positions of 
these Majorana zero modes are determined by the orientation of the Zeeman field. 
This implies that the direction of the magnetic field can be 
used to control their locations, as illustrated in Fig.~\ref{fig6}. 
These results suggest that this setup provides a practical platform 
for realizing highly tunable multifold Majorana corner modes.

In a superconductor, chiral symmetry arises from the coexistence of time-reversal 
symmetry and particle-hole symmetry. Therefore, breaking time-reversal symmetry 
with a magnetic field typically implies the absence of chiral symmetry. In the 
absence of chiral symmetry, Majorana zero modes generally follow a $Z_{2}$ classification~\cite{Schnyder2008classification,kitaev2009periodic,Ryu2010topological}. 
A consequence of this is that two spatially overlapping Majorana zero modes will hybridize, 
causing their energies to split from zero to finite values. In our system, although the 
spinful time-reversal symmetry and glide mirror symmetry are broken by the magnetic field, 
we find that their combination remains invariant. This invariance gives rise to an 
effective spinless time-reversal symmetry, with the symmetry operator given by $\mathcal{T}^{\prime}=-i\mathcal{\tilde{T}}\mathcal{\tilde{M}}_{z}=\sigma_{x}s_{x}\mathcal{K}$. 
This operator satisfies $(\mathcal{T}^{\prime})^{2}=1$ and $\mathcal{T}^{\prime}\mathcal{H}_{\rm BdG}^{\prime}(\bk)(\mathcal{T}^{\prime})^{-1}=\mathcal{H}_{\rm BdG}^{\prime}(-\bk)$,
where $\mathcal{H}_{\rm BdG}^{\prime}(\bk)=\mathcal{H}_{\rm BdG}(\bk)+\mathcal{H}_{Z}$.
The combination of this effective time-reversal symmetry and particle-hole symmetry 
results in an effective chiral symmetry, with the symmetry operator given by $\mathcal{S}^{\prime}=\tau_{x}\sigma_{x}s_{x}$.
This symmetry ensures the stability of the twofold Majorana corner modes~\cite{Yin2025,Zhu2024chiral}, 
protecting them from hybridization and energy splitting.

\section{Discussions and conclusions}\label{V}

Since even-parity pairings are ubiquitous in real materials, while odd-parity 
pairings are rare, inducing odd-parity superconductivity in carefully designed 
structures based on even-parity superconductors offers a promising alternative 
pathway to explore unique phases associated with odd-parity superconductivity. The $\pi$-junction formed 
by even-parity superconductors offers a practical pathway for inducing odd-parity 
pairings in bilayer and thin-film systems. Notably, this mechanism preserves time-reversal 
symmetry, thereby enabling the realization of time-reversal-invariant TSC phases.
In this work, we show that when a 2D DSM is sandwiched bwteen two $s$-wave or $s_{\pm}$-wave 
superconductors with a $\pi$ phase difference in their pairings, time-reversal-invariant 
first-order TSCs characterized by a pair of helical Majorana edge states can be achieved 
when the system has only a Fermi surface segment in the normal state. 
Additionally, we identify an unconventional TSC phase characterized by two pairs of helical Majorana edge states when the pairing nodes lie between two Fermi surface segments, even though the bulk topological invariants, such as the mirror Chern number and the $Z_{2}$ invariant, 
take the trivial value. The robustness of two pairs of helical Majorana edge states can be attributed to weak topological invariants 
defined along high symmetry lines within the Brillouin zone. Interestingly, applying an in-plane 
magnetic field can gap out these helical Majorana edge states, leading to the emergence of 
Majorana corner states. Due to the presence of an effective chiral symmetry, we demonstrate 
that this system can host not only isolated Majorana corner modes but also twofold Majorana corner modes.

To date, a variety of materials have been predicted to host Dirac points within their band structures 
such as black phosphorus~\cite{Kim2017DSM}, FeSe~\cite{Young2017DSM}, 
X$_3$SiTe$_6$ ($X$ = Ta, Nb)\cite{Li2018NSDSM,Sato2018NSDSM}, 
SbSSn~\cite{Jin2020DSM}, HfGe$_{0.92}$Te\cite{Chen2022DSM}, and others~\cite{Meng2022DSM}. 
Furthermore, the $\pi$ phase difference in pairings can be controlled using SQUID technology, which is well-established. 
These material candidates, combined with the advanced control over phase differences, 
provide a solid foundation for realizing our proposed systems. As a final remark, 
interaction-driven odd-parity pairing could, in principle, also arise in this metallic system,
just similar to previous studies~\cite{Deng2012TRITSC,Nakosai2012,Xu2024helical}. 
We leave the exploration of this scenario for future work.   

In conclusion,  our work demonstrates that $\pi$-junction DSMs can support a rich variety of TSCs, 
ranging from first-order to second-order TSC phases. This offers a versatile platform for 
exploring exotic topological phenomena and advancing the field of topological superconductivity.

\section*{Acknowledgements}

This work is supported by the National Natural Science Foundation of China (Grant No.12174455) 
and Guangdong Basic and Applied Basic Research Foundation (Grant No. 2023B1515040023).

\appendix

\bibliography{pi_junction}

\end{document}